\documentclass[12pt]{article}

\begin{document}
\pagestyle{empty}
\begin{flushright}
{\it Dedicated to the 70th Birthday of Prof. Yu.A. Simonov.}
\end{flushright}
\vspace*{3mm}
\begin{center}
{\Large \bf
QCD sum rules and radial excitations of light pseudoscalar and 
scalar mesons}

\vspace{0.1cm}

{\bf A.L. Kataev}\footnote{E-mail:kataev@ms2.inr.ac.ru }\\
\vspace{0.1cm}

Institute for Nuclear Research of the Russian  Academy of Sciences
,\\ Moscow \\

\end{center}
\begin{center}
{\bf ABSTRACT}
\end{center}
The calculations of masses and decay constants of the radial 
excitations of light pseudoscalar and scalar mesons within 
QCD sum rules method  are briefly reviewed. 
The predictions  are based on  the $1/N_c$-supported 
model spectra, which consist of  infinite number of infinitely narrow 
resonances, and on the assumption  that  the ground states of light 
scalar  mesons  may be considered  as  the $\overline{q}q$-bound states.
The results of the studies are compared with the existing experimental data 
and with the predictions of other 
theoretical approaches.  
\noindent 
\vspace*{0.1cm}
\noindent
\vfill\eject

\setcounter{page}{1}
\pagestyle{plain}
                
\section{Introduction}

~~~~~Despite the fact that QCD is an indisputable theory of strong interactions,
there are still the number of important open areas, where using 
QCD methods, one can arrive to different conclusions which can serve as
various alternative  descriptions of the results of concrete   
experimental studies.

The status of QCD predictions for 
 the properties of  light 
hadronic bound states  is among opened and intriguing problems 
of modern phenomenology. 
The main  puzzle is that 
the long-awaited
glueball states have not yet  well-identified candidates  
even  in the most prominent 
scalar channel. Indeed, different phenomenological works indicate that  
they can mix with low-lying scalar mesons with the mass of over 1 {\rm GeV}
(see, e.g.,  \cite{Close:2002zu}, for a review). It is known that this sector 
is rich in  different scalar resonances, like $I=0$ $\delta$ or 
$f_0(600)$, $f_0(980)$, $f_0(1370)$, $f_0(1500)$, $I=1/2$ $K^*(1430)$, 
$I=1$ $a_0(980)$ and $a_0(1450)$ (see the most recent Particle Data 
Group  report   \cite{Hagiwara:fs} ). This list is minimal and does 
not include all actual resonances, which can manifest themselves, say,  as the 
radial excitations of low-lying scalar hadronic states, systematised, 
e.g., in \cite{Anisovich:2000kx}.

Moreover, there are different points of view on the nature  of even low-lying 
scalar hadrons, such as  $a_0(980)$ particle. Indeed, 
it is described  by different authors either as the 4-quark state  
\cite{Achasov:1997ih}, $\overline{q}q$-quark mixed structure 
\cite{Gerasimov:2003tn},
or the $\overline{q}q$-bound state. The 
latter point of view is supported by 
the studies of the works \cite{Umekawa:2004js} and \cite{Narison:2000dh}.

The important logical check of the assumption   that 
scalar mesons  may  be considered as the $\overline{q}q$-bound states 
is based on the investigations of the possibilities to predict 
masses and decay constants  of their  radial excitations 
and on the comparisons of the predictions of different methods. 
It is also interesting  to study the  results of applications 
of these methods  in the pseudoscalar channel, where 
low-lying $\overline{q}q$ states are well-identified as $\pi$ 
and $K$ mesons.

Quite recently several  
approaches were developed, which have the aim to  
describe the properties  of radial excitations of light 
mesons  in 
various  channels.  Among others we can  mention  
the QCD--string--inspired methods of  the works   \cite{Badalian:2002xy},
\cite{Afonin:2004yb}, the methods of the Effective Chiral 
Lagrangians \cite{Fedorov:2003zy} and the  large-$N_c$ 
expansion--motivated  approaches (see, e.g., \cite{Afonin:2004yb}).

Let us remind that within pure $1/N_c$ expansion, originally proposed in   
\cite{'tHooft:1973jz},  all quark antiquark mesons are becoming infinitely 
narrow 
resonances. The spectrum of the theory in the large-$N_c$ limit 
consists of an infinite number  
of these  resonances, which belong to flavor nonets 
\cite{Witten:1979kh}.        
Of course, in the real world we have $N_c=3\neq \infty$ and the 
families of resonances with non-zero widths. However, the  $1/N_c$-motivated 
spectrum                     
is often used in various concrete physical considerations  in 
four-dimensional QCD. Indeed, 
the phenomenological spectra, modeled 
by the comb of infinite number of infinitely narrow resonances
was used, e.g., in   
\cite{Afonin:2004yb,Golterman:2001pj}  
and in  the older   works 
 \cite{Krasnikov:1981vw}-\cite{Gorishnii:1983zi},
which attracted definite  interest  again only recently 
(see, e.g., the comparison of the results 
obtained in  \cite{Krasnikov:1981vw}-\cite{Gorishnii:1983zi}
with the ones from  \cite{Afonin:2004yb} 
and 
\cite{Fedorov:2003zy}).

These works, like some other 
similar analyses, are based on the concept of duality, proposed
within the context of QCD in  \cite{Bramon:1972vv,Sakurai:rh}
some  time ago, 
tested in two-dimensions within $1/N_c$-expansion in   
\cite{Bradley:aq},
and studied in detail in the review of 
\cite{Shifman:2000jv}. 
The method of the QCD finite-energy sum rules (FESRs) \cite{Chetyrkin:ta}, 
which are  analogous to the dual sum rules in the 
theory of strong interactions  \cite{Logunov:1967dy},
is another important technical tool, applied  for theoretical
investigations of the properties of radial excitations of mesons,
performed in  \cite{Krasnikov:1981vw}-\cite{Gorishnii:1983zi}.

Since quite recently definite  
interest to the studies of the predictions 
for the 
spectra of radial excitations of light hadronic states was observed
which is   
motivated in part by the experimental programs of the collaboration 
COMPASS at CERN and CEBAF Jefferson Laboratory facility, we  
decided to remind some  basic theoretical  results, obtained in the beginning 
of 90's in the Theoretical Division of Institute for Nuclear Research,
concentrating on the considerations of the pseudoscalar  and scalar  
sectors.

\section{The pseudoscalar and scalar channel: preliminaries}

~~~~~Let us first introduce the (pseudo)scalar quark currents   
\begin{equation}
\partial_{\mu}J_{\mu}^{(5)}=im_q^{(+)-}\overline{q}(\gamma_5)u~~~,
\end{equation}
which are  proportional 
to  the divergence of (axial)vector currents:
\begin{equation}
J_{\mu}^{(5)}=\overline{q}\gamma_{\mu}(\gamma_5)u~~~,
\end{equation}
where $m_q^{(+)-}=m_q\pm m_u$  
are the sum and difference 
of the current quark masses with  $q=d,s$.
The two-point functions of the (pseudo)scalar quark currents can 
be defined as:
\begin{equation}
\Pi^{(P)S}(Q^2)=i(8\pi^2)\int e^{iqx}<0|\partial_{\mu}J_{\mu}^{(5)}(x) 
~\partial_{\mu}J_{\mu}^{(5)}(0)|>d^4x~~~~.
\end{equation} 
where $Q^2=-q^2$ is the Euclidean momentum transfer, and indexes 
$P$ and $S$ are labeling the pseudoscalar and scalar quark  channels.
These two-point functions  have  the following imaginary part 
\begin{equation}
\label{R}
R^{(P)S}_{\rm{\overline{MS}}}(s/\mu^2)=3(m_q^{(+)-}(\mu^2))^2s
\bigg[1+\frac{\alpha_s(\mu^2)}{\pi}
\bigg(\frac{17}{4}C_F-\frac{3}{2}C_F\rm{ln}(s/\mu^2)\bigg)+...
\bigg]~~~~~,
\label{spd}
\end{equation}
where $s$ is the Minkowskian variable
and $\mu^2$ is the normalisation point of the $\rm{\overline{MS}}$-scheme.  
Here  $C_F=(N_c^2-1)/(2N_c)$ is one of the Casimir operators 
and we retain the 1-loop massless perturbative 
QCD correction only. The result for this correction can be 
extracted from the original $\rm{\overline{MS}}$-scheme calculation of 
\cite{Becchi:1980vz}. The spectral density of Eq. (\ref{spd}) enters in 
the following Euclidean function \cite{Broadhurst:2000yc} 
\begin{equation}
\label{DS}
D^{(P)S}(Q^2)=Q^2\int_0^{\infty}\frac{R^{(P)S}_{\rm{\overline{MS}}}(s/\mu^2)}
{(s+Q^2)^3}ds~~~~~,  
\label{D}
\end{equation}
which 
obeys the renormalisation group equation 
with anomalous mass dimension term, namely  
\begin{equation}
\bigg(\frac{\partial}{\partial\rm ln(\mu^2)}+\beta(a_s)
\frac{\partial}{\partial a_s}+2\gamma_m(a_s)\frac{\partial}{\partial 
m_q^{(+)-}}\bigg)D^{(P)S}(Q^2/\mu^2)=0~~~~~~.
\end{equation}
Here  $a_s(\mu^2)=\alpha_s(\mu^2)/(4\pi)$ and $m_q^{(+)-}(\mu^2)$ are 
related to the renormalised coupling constant and quark masses, which depend 
from  
the normalisation point $\mu^2$. The QCD $\beta$-function and 
the anomalous dimension $\gamma_m$ of  quark mass  $m_q$ 
are defined as 
\begin{equation}
\beta(a_s)=\frac{d~a_s(\mu^2)}{d~{\rm{ln}~\mu^2}}=-\sum_{n\geq0}
\beta_n\bigg(\frac{\alpha_s}{4\pi}\bigg)^{n+2}~~~~,
\end{equation}
\begin{equation}
\gamma_m(a_s)=\frac{d~\rm{ln}~m_q(\mu^2)}{d~\rm{\ln}~(\mu^2)}=-\sum_{n\geq0}
\gamma_n\bigg(\frac{\alpha_s}{4\pi}\bigg)^{n+1}~~~.
\end{equation} 
In  the class of the $\rm{\overline{MS}}$-schemes the analytical 
expressions for the first two coefficients of the 
QCD $\beta$-function and the  anomalous dimension function $\gamma_m$ 
read:
\begin{eqnarray}
\beta_0&=&\bigg(\frac{11}{3}C_A-\frac{2}{3}N_F\bigg)~~, \\
\beta_1&=&\bigg(\frac{34}{3}C_A^2-2C_FN_F-\frac{10}{3}C_AN_F\bigg)
~~~,
\\
\gamma_0&=&3C_F~~~,\\
\gamma_1&=&\bigg(\frac{3}{2}C_F^2+\frac{97}{2}C_FC_A-5C_FN_F\bigg)
~~~,
\end{eqnarray}
where $C_F$ was introduced previously and $C_A=N_c$.
Within the class of gauge-independent schemes  the coefficient 
$\gamma_1$  is scheme dependent, 
while the coefficients $\beta_0$, $\beta_1$ and $\gamma_0$ 
do not depend from the choice of the subtraction scheme.
 Note that the general analytical 
4-loop  
expressions for the QCD $\beta$-function and the  
$\gamma_m$-function  were calculated  in the works of 
\cite{vanRitbergen:1997va}
and \cite{Vermaseren:1997fq} respectively. The  results from 
\cite{Vermaseren:1997fq}, expressed in terms of $N_c$,  
are  in agreement with the outcomes of independent calculation of  
\cite{Chetyrkin:1997dh}. As to the  perturbative corrections to the 
function of Eq. (\ref{D}), we will limit ourselves by the 
consideration of the following 1-loop expression: 
\begin{equation}
\label{PS}
D^{(P)S}(Q^2)=3(m_q^{(+)-}(Q^2))^2
\bigg[1+\frac{11}{3}C_F\bigg(\frac{\alpha_s(Q^2)}{\pi}\bigg)+...\bigg]~~~,
\end{equation}
where within the $\rm{\overline{MS}}$-scheme the running quark masses are 
related to 
the invariant quark masses $\hat{m_q}^{(+)-}$ as 
\begin{eqnarray}
\label{RI}
\nonumber
m_q^{(+)-}(Q^2)&=&\hat{m}_q^{(+)-}~\rm{exp}~\bigg[ \int_0^{\it{a}_s(Q^2)}
\frac{\gamma_m(x)}{\beta(x)}dx-\int_0^{2\beta_0}\frac{\gamma_0}{\beta_0x}dx
+2\frac{\gamma_0}{\beta_0}{\rm{ln}}(2\beta_0)\bigg] \\ 
&=&\hat{m}_q^{(+)-}(\beta_0\alpha_s(Q^2)/2\pi)^{\gamma_0/\beta_0}
\bigg[1+\bigg(\frac{\gamma_1}{\beta_1}-\frac{\gamma_0}{\beta_0}
\bigg)\frac{\beta_1}{\beta_0}\bigg(\frac{\alpha_s}{4\pi}\bigg)+...\bigg]~~~,
\end{eqnarray}
and  the  2-loop  expression for the QCD  
coupling constant in the $\rm{\overline{MS}}$-scheme, 
which corresponds to $N_F=3$ numbers of active flavours, 
reads 
\begin{equation}
\alpha_s(Q^2)=\frac{4\pi}{\beta_0\rm{ln}(Q^2/\Lambda^2)}\bigg[1-
\frac{\beta_1 \rm{ln~ln}(Q^2/\Lambda^2)}{\beta_0^2\rm{ln}(Q^2/\Lambda^2)}\bigg]
, ~~~  \Lambda=\Lambda_{\rm{\overline{MS}}}^{(N_F=3)}~~.
\end{equation}
Note that the  relation 
between the running and the  invariant quark masses  was 
originally chosen in  \cite{Becchi:1980vz} in the way of Eq.(\ref{RI})
to fix the following scale-dependence of the  quark running mass : 
\begin{equation}
m_q(\mu)=\hat{m}_q\bigg(\frac{1}{\rm{ln}(Q/\Lambda)}\bigg)^{\gamma_0/\beta_0}~~~.
\end{equation}

Notice, that  the application  of the $1/N_c$-expansion is used in QCD
for  the choice of model phenomenological spectral function only.
Of course, one can be more consistent, expanding  all Casimir operators $C_F$ 
and $C_A$ 
in powers of $N_c$ and keeping  the leading terms of this expansion. 
In this case one may use the following formulae: 
\begin{eqnarray}
\frac{m_q^{(+)-}(Q^2)}{m_q^{(+)-}(\mu^2)}&=&\bigg(\frac{\alpha_s(Q^2)}
{\alpha_s(\mu^2)}\bigg)^{11N_c^2/2}~~~~, \\
\frac{\alpha_s(Q^2)}{\pi}&=&\frac{12\pi}{11N_c\rm{ln}(Q^2/\Lambda^2)}~~~ .
\end{eqnarray}
However, we will avoid applications of these formulae in the  concrete 
considerations, which will be discussed below.
 
It should be stressed, that the convention of choosing the renormalisation 
scale of running quark masses at $Q^2=1~{\rm{GeV}^2}$ moved in our times to 
$2 ~\rm {GeV^2}$. Moreover, since partly unknown QCD corrections 
of order $(\alpha_s/\pi)^4$ to Eq. (\ref{R}) and Eq. (\ref{PS})
(the available results of the 
total calculations of this term, which are  now continuing \footnote{Private 
communications by P. A. Baikov and K. G. Chetyrkin are gratefully acknowledged
}, 
see in  \cite{Broadhurst:2000yc} and  \cite{Baikov:2001aa})
may affect the precision of the determinations of light quark masses from the 
scalar and pseudoscalar quark channels, we will not  present here   
any concrete results both for the running and invariant quark masses. 

Before proceeding to the main part of this work let us emphasise the 
essential role of the $<m_q\overline{q}q>$ and 
$<(\beta(\alpha_s)/\alpha_s)G_{\mu\nu}^aG_{\mu\nu}^a>$  condensates 
\cite{Shifman:bx} in the description of the properties of the ground 
states of hadrons. Indeed, together with instanton effects 
(see, e.g.,  \cite{Novikov:xj,Nason:1994xw}), these nonperturbative 
contributions should be  mostly 
important for the  calculations  of the hadronic ground states 
masses and decay width coupling constant 
using the OPE technique and   the Borel 
sum rules approach  \cite{Shifman:bx}. 
Note, that the infrared renormalon 
calculus, rediscovered in QCD in  \cite{Zakharov:1992bx}, supports the
importantce of consideration of the condensates with dimension $d\geq 4$. 
Moreover, this approach favors the application of the 
dispersion  relation of Eq. (\ref{D}) in the scalar and pseudoscalar channels.
Due to the theoretical arguments,  given in \cite{Broadhurst:2000yc},
the renormalon calculus indicate  that 
the ill-defined dispersion relation 
\begin{equation}
\overline{D}^{P(S)}(Q^2)=Q^2\int_0^{\infty}\frac{R_{\rm\overline{MS}}^
{P(S)}(s)}{s(s+Q^2)^2}ds~~~,
\end{equation}    
contains  $(\Lambda^2/Q^2)$-correction,
which is not consistent with the general structure of the standard  massless 
operator-product expansion (OPE)  technique. This fact 
indicates 
that  
in the process of the concrete phenomenological studies
it is more consistent to consider  
the $D$-function 
of Eq. (\ref{DS})  
\cite{Broadhurst:2000yc}
(in fact it is proportional to the  double-differentiated dispersion  
relation, originally defined and used  in  \cite{Becchi:1980vz}).
Indeed, like in the vector 
channel, the standard  OPE expansion  of Eq. (\ref{DS}) is starting from 
the terms   
of order $O(\Lambda^4/Q^4)$. Thus, following the scalar-channel consideration 
of \cite{Broadhurst:2000yc}, we conclude that in the 
$\rm{\overline{MS}}$-scheme the infrared renormalon (IRR) calculus 
is rather useful tool for the investigation of 
the general structure of the standard   OPE formalism.
Note, however, that the concept of the IRR-contributions is scheme-dependent.
In the schemes with the frozen coupling constant
(see, e.g.,  \cite{Krasnikov:1996jq}-\cite{Simonov:2001di})
the IRR contributions are absent. The question of the existence in this 
case of the ultraviolet renormalon contributions seems to be opened.
Another problem, which we are going to concentrate in this presentation, is 
related to the possibility to estimate the properties of radial excitations,
namely their masses and decay width coupling constants, using the QCD sum 
rules method and the duality approach. In the next sections we will 
consentrate on  the  
the analysis 
of this problem in the pseudoscalar and scalar quark channels.  

\section{Radial excitations of light pseudoscalar mesons}

~~~~~In the works of  \cite{Kataev:1982xu,Gorishnii:1983zi}
the properties of radial excitations of light pseudoscalar mesons were 
studied  
with the help of  the following  FESR  
\begin{equation} 
\label{moments}
M_k^{\rm{th}}=\int_{s_{n-1}}^{s_n}R^{P}_{\rm{th}}(s)s^k ds = M_k^{\rm{ph}}=
\int_{s_{n-1}}^{s_n}R^{P}_{\rm{ph}}(s)s^k ds~~~~,
\end{equation}
where $s_n$ are the duality intervals, which will be defined 
below, the  theoretical ($\rm{th}$)  spectral 
density of FESR
is calculated in four dimensional 
QCD, while the phenomenological ($\rm{ph}$) model for the spectral 
function of FESR    is chosen 
in the form of $1/N_c$-motivated model 
\begin{equation} 
R^P_{\rm{ph}}(s)=\sum_{l=1}^{\infty}2f_P^2(l)m_P^4(l)\delta(s-m_P^2(l))~~~~.
\end{equation}
Index $P$ labels the sets of masses $m_P(l)$  and decay constants 
$f_P(l)$   
of the radial excitations 
of light pseudoscalar mesons, namely $\pi$ amd $K$-mesons,
which are considered as massless particles.

Neglecting the slight $\alpha_s$-dependence, 
which comes from the leading-order terms 
of Eq. (4), the authors of  \cite{Kataev:1982xu,Gorishnii:1983zi}
obtained the following sum-rule :
\begin{equation}
\frac{M_0^{\rm{th}}(s_n)}{M_{-1}^{\rm{th}}(s_n)}=\frac{1}{2}(s_{n}+s_{n-1})=  
\frac{M_0^{\rm{ph}}(s_n)}{M_{-1}^{\rm{ph}}(s_n)}= m_P^{2}(n)~~~~.
\end{equation}
As the next step the bounds of integration in Eq. (\ref{moments}) where 
chosen as 
\begin{eqnarray}
s_n&=&\frac{1}{2}\bigg[m_P^2(n)+m_P^2(n+1)\bigg]~~~~ , \\
s_0&=&\frac{m_{P}(1)}{2}~~~~,
\end{eqnarray}
where $m_P(1)$ is  the mass of the first radial excitaions 
of $\pi$ mesons, namely the $\pi^{'}$  state.
The choice of these duality intervals 
is supported by the studies of possibilities 
to combine the $1/N_c$-motivated spectrum with the duality approach in two 
and 
four dimensions \cite{Bradley:aq}. 

As the results of iterative solution of the system of Eqs. (22)-(24) the 
following mass formula for the radial  excitations of the  $\pi$ meson 
was obtained 
\cite{Kataev:1982xu,Gorishnii:1983zi}:
\begin{equation}
\label{pir}
m_{\pi}^2(l)=m_{\pi^{ '}}^2l~~~ ,~~~ l\geq 1~~~.
\end{equation}
In the work  \cite{Gorishnii:1983zi} these considerations were generalised to 
the 
case of the $K$ meson radial excitations and  the identical 
expression for the mass spectrum 
\begin{equation}  
m_{K}^2(l)=m_{K^{ '}}^2l~~~ ,~~~ l\geq 1~~~~
\end{equation}
was derived.

It should be stressed that the $\pi^{'}$ meson was observed by several 
experimental collaborations (see \cite{Hagiwara:fs}). In the work  
\cite{Gorishnii:1983zi} the result $m_{\pi^{'}}=1240~{\rm{MeV}}$,
obtained at Protvino accelerator
by Dubna--Milan--Bologna Collaboration 
 \cite{Bellini:ec}, was used.  Substituting it into   Eq. (\ref{pir}), 
it is easy to get the  following predictions $m_{\pi}(2)=1753~{\rm{MeV}}$, 
 $m_{\pi}(3)=2148~{\rm{MeV}}$,   $m_{\pi}(4)=2480~{\rm{MeV}}$ 
\cite{Gorishnii:1983zi}. These numbers  are in good agreement with the linear 
trajectories, obtained in  \cite{Anisovich:2000kx}, and with the results 
of the recent OPE-based analysis of the work  \cite{Afonin:2004yb}. 
Note  also, that 
the application of the Effective Chiral Lagrangians gives 
$m_{\pi}(2)=1.98~GeV$ \cite{Fedorov:2003zy},
 while the experimental number from  \cite{Bellini:ec}
is  $m_{\pi}(2)=1.77\pm 0.03$ ${\rm{GeV}}$. The inclusion of the 
$\pi$-meson radial excitations in the studies of the QCD sum rule model for 
the pion wave-function gave  the following prediction:  
 $m_{\pi}(2)=2.05\pm 0.15$ ${\rm{GeV}}$ \cite{Radyushkin:1994xv}. 
It  is consistent with the result of  \cite{Fedorov:2003zy},
but is slightly higher than the  prediction from \cite{Gorishnii:1983zi},
which is in surprisingly good agreement with the experimental result 
of   \cite{Bellini:ec}.

Fixing  now the experimental value of $m_{K^{'}}$=1.46 GeV 
\cite{Hagiwara:fs} as the input parameter, it is possible to 
obtain the predictions 
for the masses of higher radial excitations of $K$ meson 
\cite{Gorishnii:1983zi}, namely $m_{K}(2)=2.06~{\rm {GeV}}$ and 
$m_{K}(3)=2.53~{\rm{GeV}}$. The prediction for the mass 
of the second radial excitation is consistent with the result 
$m_{K}(2)=2.1~{\rm {GeV}}$ obtained in  \cite{Fedorov:2003zy}, which is 
slightly higher than the experimental result  $m_{K}(2)=1.86~{\rm {GeV}}$ 
\cite{Hagiwara:fs}. The comparison  of the results 
of  \cite{Gorishnii:1983zi} in the $K$-meson channel 
with the results of other theoretical studies 
are  really welcomed.

As the next step the FESR model for the decay   
constants of the radial excitations of light pseudoscalar mesons was estimated
\cite{Kataev:1982xu,Gorishnii:1983zi}. 
Reminding that the duality interval of the ground state for the 
pseudoscalar particles can be defined as  $s_0=m_{P}^2(1)/2$,
taking the ratio of the 
following FESRs
\begin{equation}
\frac{M_2^{\rm{th}}(s_n)}{M_2^{\rm{th}}(s_o)}=\frac{s_{n+1}^2-s_n^2}{s_0^2}=
\frac{{f_P}^2(n)m_P^{4}(n)}{f_{P}^2m_P^4}
\end{equation}
and supplementing it with Eq. (23), it is possible to get  the following 
``linear dual model'' for the coupling constants of radial excitations 
of the light pseudoscalar mesons \cite{Kataev:1982xu,Gorishnii:1983zi}:
\begin{equation}
\label{cc}
f_P(l)=2\sqrt{2}\frac{m_P^2}{m_P(1)m_P(l)}f_P,~~~ l\geq 1~~~ .
\end{equation}
Taking  into account the concrete values for   
$m_{\pi}$, $m_{K}$, the values of the  
decay constants $f_{\pi}$ and $f_K$ and the 
expressions for the masses of radial excitations of the pseudoscalar  
mesons $m_{\pi}(l)$ and $m_{K}(l)$, one can get the  FESR-inspired 
estimates for the decay constants of radial excitations of 
$\pi$ and $K$ mesons.
It will be interesting to study the possible numerical uncertainties 
of this model using 
other approaches.
 
To conclude this section it is also worth  mentioning  that 
the similar ``linear dual spectrum'' in the vector channel 
\cite{Krasnikov:1981vw} can be also obtained within Veneziano model  
 \cite{Veneziano:yb}.

\section{Radial excitations of light scalar mesons}

~~~~~We are now ready to discuss the most intriguing part of the work 
 \cite{Gorishnii:1983zi}, which is devoted to the derivation 
of ``linear dual spectra'' in the light scalar meson channel, 
whose ground state representatives $a_0(980)$ and $K^*(1430)$ 
will be considered as the massive $\overline{q}q$-bound states.
The only difference with the discussions presented in Sect.3 is related 
to the redefinition of the ground-state duality interval from  
$(s_0)_P=m_P^2(1)/2$ to $(s_0)_S=3m_S^2/2$, where $m_S$ are the 
masses of $a_0(980)$ and $K^*(1430)$ light scalar mesons. 
This definition of $(s_0)_S$ is coming from the
following ratio of the FESRs:
\begin{equation}
\frac{M_1^{\rm{th}}(s_0)}{M_0^{\rm{th}}(s_0)} =\frac{2(s_0)_{S}}{3}=m_S^2~~~.
\end{equation}
Combining this new value of the duality interval with Eq. (22) 
and Eq. (23)  
the authors of  \cite{Gorishnii:1983zi} obtained 
the following scalar analog of the ``pseudoscalar linear  dual 
model''
derived in the previous 
section 
 \cite{Gorishnii:1983zi}:
\begin{eqnarray}
m_S(n)&=&(n+1)m_S^2~~~,\\
f_S^2(n) &=& \frac{1}{n+1}f_S^2~~~~.
\end{eqnarray}
Thus, assuming the $\overline{q}q$-structure of $a_0(980)$ meson, 
we  expected that the masses of  its radial 
excitations may be estimated as  
$a_0(1)=1380~{\rm MeV}$, $a_0(2)=1697~{\rm MeV}$,
$a_0(3)=1960~ {\rm MeV}$. Note, that the work  \cite{Gorishnii:1983zi}
was the first one, where the possibility of the existence of extra 
light scalar resonance near $a_0(1)=1.4~ {\rm GeV}$ was predicted. 
It is known now 
that there is $a_0(1450)$ meson in nature \cite{Hagiwara:fs}. 
So, it may treated as the  possible candidate for the 
first radial excitation 
of $a_0(980)$ meson. Another pleasant feature of the derived 
in  \cite{Gorishnii:1983zi} ``linear dual spectrum'' 
for the  $a_0$-meson excitations is that it turns out to be in 
satisfactory agreement with the results from \cite{Afonin:2004yb}.
As to the strange light scalar particle, namely $K_0^*(1430)$ meson,
within described above approach 
its possible radial excitations should   have the following masses:
$m_{K_0^*}(1)=2022~{\rm MeV}$ and  $m_{K_0^*}(2)=2477~{\rm MeV}$.
Note, that the experimental data indicate the existence of 
$K_0^*(1950)$ meson, which following discussed above  
classification, can be considered  
as the candidate for the first radial excitation of  $K_0^*(1430)$ meson.
This meson may be the good candidate for the nonet partner of $a_0(1450)$ meson. 
However,  to get better understanding on the 
structure of the scalar nonets and the nature of both 
$a_0(980)$ and $a_0(1450)$ mesons 
it is rather important to continue studies of 
the  classification of hadrons  in the 
light scalar 
sector  using various  approaches.

\section{Conclusion}

~~~~In this work the applications of the  QCD FESR approach 
for the 
derivation of ``linear dual spectra'' of radial excitations in the 
pseudoscalar and scalar quark antiquark channel were reminded. 
In the process of considerations both 
nonperturbative and, 
probably more important in the investigation of this problem, perturbative 
QCD effects, were neglected.  
It is worth to emphasise that  higher-order 
perturbative QCD corrections for the massless
two-point function of 
pseudoscalar and scalar quark currents, calculated in the works of  
\cite{Gorishnii:1990zu,Chetyrkin:1996sr}, are more important, than 
the ones for the two-point function of the  vector channel, calculated in the 
works      
\cite{Chetyrkin:bj,Gorishnii:1990vf}. In view of this it can be of interest 
to take these calculated corrections  into account in the studies of the 
properties 
of radial excitations of light mesons, based on the   OPE approach.

I would like to thank Prof. Yu. A. Simonov for stimulating me to remind  
the results of the works 
\cite{Kataev:1982xu,Gorishnii:1983zi}, made in the beginning of 90's. 
The work on this mini-review was done within the framework of the RFBR 
Grants N 02-01-00601, 03-02-17047, and 03-02-17177.


\begin{thebibliography}{99}
\bibitem{Close:2002zu}
F.~E.~Close and N.~A.~Tornqvist,
J.\ Phys.\ G {\bf 28},  R249 (2002)
[hep-ph/0204205].
\bibitem{Hagiwara:fs}
K.~Hagiwara {\it et al.}  (Particle Data Group Collab.),
Phys.\ Rev.\ D {\bf 66},  010001 (2002)
\bibitem{Anisovich:2000kx}
A.~V.~Anisovich, V.~V.~Anisovich, and A.~V.~Sarantsev,
Phys.\ Rev.\ D {\bf 62},  051502 (2000)
[hep-ph/0003113].
\bibitem{Achasov:1997ih}
N.~N.~Achasov and V.~V.~Gubin,
Phys.\ Rev.\ D {\bf 56},  4084 (1997)
[hep-ph/9703367].
\bibitem{Gerasimov:2003tn}
S.~B.~Gerasimov,
{\it in  12th International  Conference on Selected Problems of 
Modern Physics (Blokhintsev 03), Dubna, Russia, 8-11 June 2003},
hep-ph/0311080.
\bibitem{Umekawa:2004js}
T.~Umekawa, K.~Naito, M.~Oka and M.~Takizawa,
hep-ph/0403032.
\bibitem{Narison:2000dh}
S.~Narison,
Nucl.\ Phys.\ Proc.\ Suppl.\  {\bf 96}, 244  (2001)
[hep-ph/0012235].
\bibitem{Badalian:2002xy}
A.~M.~Badalian, B.~L.~G.~Bakker, and Y.~A.~Simonov,
Phys.\ Rev.\ D {\bf 66},  034026 (2002)
[hep-ph/0204088].
\bibitem{Afonin:2004yb}
S.~S.~Afonin, A.~A.~Andrianov, V.~A.~Andrianov, and D.~Espriu,
JHEP {\bf 0404}, 039  (2004) 
[hep-ph/0403268].
\bibitem{Fedorov:2003zy}
S.~M.~Fedorov and Y.~A.~Simonov,
JETP Lett.\  {\bf 78},  57 (2003)
[Pisma Zh.\ Eksp.\ Teor.\ Fiz.\  {\bf 78},  67 (2003) ]
[hep-ph/0306216].
\bibitem{'tHooft:1973jz}
G.~'t Hooft,
Nucl.\ Phys.\ B {\bf 72},  461 (1974).
\bibitem{Witten:1979kh}
E.~Witten,
Nucl.\ Phys.\ B {\bf 160},  57 (1979).
\bibitem{Golterman:2001pj}
M.~Golterman, S.~Peris, B.~Phily, and E.~De Rafael,
JHEP {\bf 0201},   024 (2002)
[hep-ph/0112042].
\bibitem{Krasnikov:1981vw}
N.~V.~Krasnikov and A.~A.~Pivovarov,
Phys.\ Lett.\ B {\bf 112},  (1982) 397.
\bibitem{Kataev:1982xu}
A.~L.~Kataev, N.~V.~Krasnikov, and A.~A.~Pivovarov,
Phys.\ Lett.\ B {\bf 123}, (1983) 93.
\bibitem{Gorishnii:1983zi}
S.~G.~Gorishny, A.~L.~Kataev, and S.~A.~Larin,
Phys.\ Lett.\ B {\bf 135}, 457 (1984).
\bibitem{Bramon:1972vv}
A.~Bramon, E.~Etim, and M.~Greco,
Phys.\ Lett.\ B {\bf 41}, 609 (1972).
\bibitem{Sakurai:rh}
J.~J.~Sakurai,
Phys.\ Lett.\ B {\bf 46}, 207 (1973).
\bibitem{Bradley:aq}
A.~Bradley, C.~S.~Langensiepen, and G.~Shaw
Phys.\ Lett.\ B {\bf 102}, 180 (1981).
\bibitem{Shifman:2000jv}
M.~A.~Shifman, {\it in Boris Ioffe Festschrift ``At the Frontier of Particle 
Physics/Handbook of QCD'',} (World Scientific, Singapore, 2001), Vol.3, p.1447 
[hep-ph/0009131].
\bibitem{Chetyrkin:ta}
K.~G.~Chetyrkin, N.~V.~Krasnikov, and A.~N.~Tavkhelidze,
Phys.\ Lett.\ B {\bf 76},  83 (1978).
\bibitem{Logunov:1967dy}
A.~A.~Logunov, L.~D.~Soloviev, and A.~N.~Tavkhelidze,
Phys.\ Lett.\ B {\bf 24} 181 (1967).
\bibitem{Becchi:1980vz}
C.~Becchi, S.~Narison, E.~De Rafael, and F.~J.~Yndurain,
Z.\ Phys.\ C {\bf 8}, 335 (1981).
\bibitem{Broadhurst:2000yc}
D.~J.~Broadhurst, A.~L.~Kataev, and C.~J.~Maxwell,
Nucl.\ Phys.\ B {\bf 592}, 247 (2001)
[hep-ph/0007152].
\bibitem{vanRitbergen:1997va}
T.~van Ritbergen, J.~A.~M.~Vermaseren, and S.~A.~Larin,
Phys.\ Lett.\ B {\bf 400}, 379  (1997)
[hep-ph/9701390].
\bibitem{Vermaseren:1997fq}
J.~A.~M.~Vermaseren, S.~A.~Larin, and T.~van Ritbergen,
Phys.\ Lett.\ B {\bf 405}, 327  (1997)
[hep-ph/9703284].
\bibitem{Chetyrkin:1997dh}
K.~G.~Chetyrkin,
Phys.\ Lett.\ B {\bf 404}, 161 (1997) 
[hep-ph/9703278].
\bibitem{Baikov:2001aa}
P.~A.~Baikov, K.~G.~Chetyrkin, and J.~H.~Kuhn,
Phys.\ Rev.\ Lett.\  {\bf 88},  012001 (2002)
[hep-ph/0108197].
\bibitem{Shifman:bx}
M.~A.~Shifman, A.~I.~Vainshtein, and V.~I.~Zakharov,
Nucl.\ Phys.\ B {\bf 147}, 385 (1979).
\bibitem{Novikov:xj}
V.~A.~Novikov, M.~A.~Shifman, A.~I.~Vainshtein, and V.~I.~Zakharov,
Nucl.\ Phys.\ B {\bf 191}, 301 (1981).
\bibitem{Nason:1994xw}
P.~Nason and M.~Palassini,
Nucl.\ Phys.\ B {\bf 444}, 310 (1995)
[hep-ph/9411246].
\bibitem{Zakharov:1992bx}
V.~I.~Zakharov,
Nucl.\ Phys.\ B {\bf 385}, 452 (1992).
\bibitem{Krasnikov:1996jq}
N.~V.~Krasnikov and A.~A.~Pivovarov,
Mod.\ Phys.\ Lett.\ A {\bf 11}, 835 (1996) 
[hep-ph/9602272].
\bibitem{Shirkov:1997wi}
D.~V.~Shirkov and I.~L.~Solovtsov,
Phys.\ Rev.\ Lett.\  {\bf 79}, 1209  (1997) 
[hep-ph/9704333].
\bibitem{Simonov:2001di}
Y.~A.~Simonov,
Phys.\ Atom.\ Nucl.\  {\bf 65}, 135  (2002) 
[Yad.\ Fiz.\  {\bf 65}, 140  (2002)]
[hep-ph/0109081].
\bibitem{Bellini:ec}
G.~Bellini {\it et al.},
Phys.\ Rev.\ Lett.\  {\bf 48},  1697 (1982).
\bibitem{Radyushkin:1994xv}
A.~V.~Radyushkin,
{\it in ``Continuous advances of QCD'', Minneapolis 1994}, (River Edge, 
World Scientific, 1994) p.238
[hep-ph/9406237].
\bibitem{Veneziano:yb}
G.~Veneziano,
Nuovo Cimento \ A {\bf 57}, 190 (1968).
\bibitem{Gorishnii:1990zu}
S.~G.~Gorishny, A.~L.~Kataev, S.~A.~Larin, and L.~R.~Surguladze,
Mod.\ Phys.\ Lett.\ A {\bf 5}, 2703 (1990);\\
S.~G.~Gorishny, A.~L.~Kataev, S.~A.~Larin, and L.~R.~Surguladze,
Phys.\ Rev.\ D {\bf 43}  1633 (1991).
\bibitem{Chetyrkin:1996sr}
K.~G.~Chetyrkin,
Phys.\ Lett.\ B {\bf 390}, 309  (1997) 
[hep-ph/9608318].
\bibitem{Chetyrkin:bj}
K.~G.~Chetyrkin, A.~L.~Kataev, and F.~V.~Tkachov,
Phys.\ Lett.\ B {\bf 85}, 277  (1979);\\
M.~Dine and J.~R.~Sapirstein,
Phys.\ Rev.\ Lett.\  {\bf 43}, 68 (1979);\\
W.~Celmaster and R.~J.~Gonsalves,
Phys.\ Rev.\ Lett.\  {\bf 44}, 560 (1980).
\bibitem{Gorishnii:1990vf}
S.~G.~Gorishny, A.~L.~Kataev, and S.~A.~Larin,
Phys.\ Lett.\ B {\bf 259}, 144 (1991);\\
L.~R.~Surguladze and M.~A.~Samuel,
Phys.\ Rev.\ Lett.\  {\bf 66}, 560  (1991)
[Erratum-ibid.\  {\bf 66}, 2416 (1991)];\\
K.~G.~Chetyrkin,
Phys.\ Lett.\ B {\bf 391}, 402 (1997)
[hep-ph/9608480].



\end{thebibliography}
\end{document}